\documentclass[sn-basic]{sn-jnl}
\usepackage{graphicx}
\usepackage{amsmath}
\usepackage{amssymb}

\jyear{2021}%

\theoremstyle{thmstyleone}%

\theoremstyle{thmstyletwo}%

\theoremstyle{thmstylethree}%

\begin{document}

\title[Anisotropic Approach: Compact star as generalized Model]{Anisotropic Approach: Compact star as generalized Model}

\author[1]{\fnm{B. S.} \sur{Ratanpal}}\email{bharatratanpal@gmail.com}
\author*[2]{\fnm{Rinkal} \sur{Patel}}\email{rinkalpatel22@gmail.com}

\equalcont{These authors contributed equally to this work.}

\affil[1]{\orgdiv{Department of Applied Mathematics}, \orgname{The Maharaja Sayajirao University of Baroda}, \orgaddress{\street{Faculty of Technology \& Engineering}, \city{Vadodara}, \postcode{390 001}, \state{Gujarat}, \country{India}}}
\affil*[2]{\orgdiv{Department of Applied Science \& Humanities}, \orgname{Parul University}, \orgaddress{\street{Limda}, \city{Vadodara}, \postcode{391 760}, \state{Gujarat}, \country{India}}}



\abstract{	\noindent  We studied a new class of interior solutions that are singularity free and useful for describing anisotropic compact star objects with spherically symmetric matter distribution. We have considered metric potential  selecting $ B^{2}_{0}(r)=\frac{1}{(1-\frac{r^2}{R^2})^n} $,where $ n>2 $. The various physical characteristics of the model are specifically examined for the pulsar  PSRJ1903+327 with its current estimated data. According to analysis, every physical need for a physically admissible star is satisfied and all features are acceptable. Further the stability of the model has been examined. Numerous physical characteristics are also highlighted in a graphical form.}

\keywords{singularity free, Exact solutions, anisotropic compact star}



\maketitle

\section{Introduction}\label{sec1}

	Strange stars and neutron stars are an interesting topic for researcher in astrophysics. The research into relativistic stellar structure has been ongoing for more than a century, ever since \cite{schwarzschild1916sitz} discovered the vacuum exterior solution. When a massive star explodes as a supernova, the phase transition between hadronic and strange quark matter may occure at a density higher than the nuclear density given by \cite{bodmer1971collapsed}. The pressure inside these compact stars can be decomposed into two parts: the radial pressure $ p_{r}  $ and transverse pressure $ p_{\perp},  $ where  $ p_{\perp}  $ is in the direction perpendicular to $ p_{r}  $ with the anisotropic factor  $ \Delta =  p_{\perp} -  p_{r} $. According to \cite{geng2021repeating} theory, the intriguingly repeating fast radio bursts (FRBs) are caused by irregular fractional collapses of a strange star's crust caused by filling it with accretion materials from its low-mass companion.The nature of the anisotropic factor depends on the existence of a solid core, the pressure of a type p-superfluid, a phase transition, a rotation, a magnetic field, a mixture of two fluids, etc. According to \cite{kippenhahn1990stellar}, anisotropy in relativistic stars may be caused by the presence of a solid core or type 3A superfluid. \cite{herrera1997local} studied about local anisotropy in a self gravitating systems. According to \cite{weber1999pulsars}, strong magnetic fields can also thought to be a source of anisotropic pressure inside a compact object. \cite{dev2002anisotropic}, \cite{dev2003anisotropic} and \cite{gleiser2004anistropic} have also discussed a model of an anisotropic star by assuming a special type of matter density.

The anisotropic model was obtained by considering the equation of state of the MIT bag model and a specific ansatz for the metric function $ g_{rr}, $ which was proposed by \cite{finch1989realistic}. An anisotropic strange star has been derived by \cite{rahaman2012strange} in Krori–Barua spacetime. A relativistic stellar model with a quadratic equation of state was proposed by \cite{sharma2013relativistic} in Finch–Skea spacetime. The earlier work was generalized to modify Finch-Skea spacetime by \cite{pandya2015modified}. \cite{bhar2015strange} obtained a new model of an anisotropic superdense star that allows conformal motions in the presence of a quintessence field, which is characterized by the parameter $ w_{q} $ with $ - 1 < w_{q} < - \frac{1}{3} $. The model has been developed based on the ansatz from \cite{vaidya1982exact}.

Numerous excellent efforts have been made on the precise solution of Einstein's field equations. 
\cite{maurya2015anisotropic} studied that the fluid solution is ideal in the Durgapal and Fuloria analogue. \cite{maurya2018role} explored a compact spherically symmetric relativistic body with anisotropic particle pressure while accounting for a spatial metric potential of the kind suggested by Korkina and Orlyanskii in order to solve the Einstein field equations. Eight alternative cases of generalised Buchdahl dimensionless parameter K were taken into consideration by \cite{maurya2019anisotropic}, who uniformly investigated each scenario. \cite{gupta2011class} investigated a class of charged superdense star analogues by utilizing a certain electric field. As suggested by \cite{maurya2017relativistic} provided novel anisotropic models for Buchdahl type ideal fluid solutions with metric potentials $ e^{\lambda} $  and $ e^{\nu} $ that has monotonically increasing functions. The effects of the pressure anisotropy on Buchdahl-type relativistic compact stars were discussed by \cite{maurya2019effect}. Three new exact solutions to the charged fluid  sphere of the Einstein's field equations were found by \cite{maurya2015three}, and they fit Ivanov's classification. Two new precise solutions to the Einstein's field equations for the ideal fluid distribution were developed using Lake's algorithm, in an investigation by \cite{maurya2015two}. \cite{dayanandan2016anisotropic} conducted a thorough analysis of the Matese and Whitman general relativity solution to assess the stability of anisotropic compact star models. Later, \cite{dayanandan2017modeling} was established charged compact star model for an anisotropic fluid distribution.  In order to create relativistic, anisotropic acceptable compact structures, \cite{tello2020class} developed the deformation function f(r) using the class I technique with gravitational decoupling. The minimally deformed Class I generator was used by \cite{maurya2021mgd} to illustrate an anisotropic solution of the Einstein field equations.

	There have been many recent investigations which include presence of charge and anisotropy in the stellar interior. A singularity free model of anisotroipic compact star was proposed by \cite{bhar2016compact} for utilizing Matese and Whitman mass function. By choosing a suitable form of radial pressure, a model of charged compact star was reported by \cite{bhar2015dark}. A study on the effect of anisotropy under Finch-Skea geometry was given by \cite{das2020study}. \cite{maharaj2012regular} presented regular models for charged anisotropic stellar bodies. \cite{sharma2017anisotropic} presented that a specific class of solutions can be used as an ‘anisotropic switch’ to examine the impact of anisotropy using Finch and Skea stellar model. \cite{thirukkanesh2018anisotropic} provides a new algorithm to generate anisotropic analogues of a large family of well-known solutions describing self-gravitating systems in fluid. The linear equation of state consistent with quark stars for charged anisotropic models was described by \cite{sunzu2014charged}. \cite{bhar2021finch} proposed a compact stellar model in presence of pressure anisotropy in modified Finch Skea spacetime using the ansatz $ B^{2}_{0}(r)=(1+\frac{r^2}{R^2})^n $.

The present paper deals with the study of generalized model by selecting anstaz  $ B^{2}_{0}(r)=\frac{1}{(1-\frac{r^2}{R^2})^n} $, where $ n>2 $. In section 2, it contains the Einstein field equations. In section 3, new model using particular anstaz is derived. In section 4, interior solution is matched to the exterior schwarzschild line element. Physical analysis of the model is described in section 5. Section 6, contains discussion.

\section{Einstein field equations}
For determining the structure of compact and massive stars, we use Einstein’s field equations
\begin{equation}\label{1}
	R_{\alpha\beta} - \frac{1}{2}R g_{\alpha\beta}= \frac{8\pi G}{C^4}T_{\alpha\beta}.
\end{equation}
Ricci tensor and energy-momentum tensor respectively denoted as $ R_{\alpha\beta} $	 and $ T_{\alpha\beta} $, R is Ricci scalar and  $ g_{\alpha\beta} $ is metric tensor. The universal gravitational constant and speed of the light denoted as $ G $ and $ C $ respectively.
We write the Schwarzschild coordinates for a 4-D spacetime with spherically symmetrical, the line element describing the interior space-time
$ x^{0}=t,x^{1}=r ,x^{2}=\theta,x^{3}=\phi $ as
\begin{equation}\label{e1}
	ds^2 = -A^{2}_{0}(r) dt^2 + B^{2}_{0}(r) dr^2 + r^2(d\theta^2 +sin^2 \theta d\phi^2),
\end{equation}
\\ where $ A_{0}(r) $ and $ B_{0}(r), $ the gravitational potentials are yet to
be determined. The energy momentum tensor for anisotropic fluid distribution is given by
\begin{equation}
	T^{\alpha}_{\beta}=(\rho+p_{r}) u^{\alpha} u_{\beta}+p_{t}g^{\alpha}_{\beta}+(p_{r}-p_{t})\nu^{\alpha}\nu_{\beta},
\end{equation}
\\with $ u^{\alpha}u_{\beta}=-\nu^{\alpha}\nu_{\beta} = 1 $ and $ u^{\alpha}\nu_{\beta}=0. $ Here the vector $ \nu^{\alpha} $ is the space-like vector and $ u_{\alpha} $ is the fluid 4-velocity
and it is orthogonal to $ \nu^{\alpha} $, $\rho $ represents the energy-density, $ p_{r}$ and $ p_{t} $ the fluid's radial and transverse pressures respectively.
\\ The Einstein field equations governing the system is then obtained as (we set $ G = c = 1 $ )
\begin{equation}\label{e2}
	8\pi\rho=\left[\frac{1}{r^2} -\frac{1}{r^2 B^{2}_{0}} + \frac{2B^{'}_{0}}{rB^{3}_{0}}\right],
\end{equation}
\begin{equation}\label{e3}
	8\pi p_{r} =\left[\frac{-1}{r^2} -\frac{1}{r^2 B^{2}_{0}} + \frac{2.A^{'}_{0}}{rA_{0}B^{2}_{0}}\right],
\end{equation}
\begin{equation}\label{e4}
	8\pi p_{\perp} =\left[\frac{A_{0}^{''}}{A_{0}B^{2}_{0}} + \frac{A_{0}^{'}}{rA_{0}B^{2}_{0}}-\frac{B_{0}^{'}}{r B^{3}_{0}} - \frac{A^{'}_{0}B^{'}_{0}}{A_{0}B^{3}_{0}}\right].
\end{equation}
In equations (\ref{e2})-(\ref{e4}), a ‘prime’ denotes differentiation with respect
to r.

Making use of equation (\ref{e3}) and (\ref{e4}),we define the anisotropy as
\begin{equation}\label{e5}
	\Delta(r)=8\pi(p_{\perp}-p_{r})
	=\left[ \frac{A_{0}^{''}}{A_{0}B^{2}_{0}} - \frac{A_{0}^{'}}{rA_{0}B^{2}_{0}}-\frac{B_{0}^{'}}{r B^{3}_{0}} - \frac{A^{'}_{0}B^{'}_{0}}{A_{0}B^{3}_{0}}-\frac{1}{r^2 B^{2}_{0}}+\frac{1}{r^2} \right],
\end{equation}
which must be zero at center $ r=0 $ of steller object.
\section{Generating Model}
For solving the system (\ref{e2})-(\ref{e4}), we have three equations with five unknowns $ (\rho,p_{r},p_{\perp},A_{0}(r),B_{0}(r)) .$
We are free to select any two of them to complete this system. As a result there are 10 possible ways to choose any two unknowns. According to studies \cite{sharma2013relativistic}, \cite{bhar2016anisotropic}, \cite{bhar2016new}  select $ B^{2} $ and $ p_{r} $, \cite{bhar2015dark} choose $ \rho $ along with $ p_{r} $, \cite{murad2015some} and \cite{thirukkanesh2018anisotropic} select $ A^2 $ with $ \Delta $ to model various compact stars. However, a very well-liked method is to select $ B^2 $ and an equation of state (EoS), which is a relation between matter density and radial pressure $ p_{r}.$ Numerous articles in this direction have been published like \cite{sunzu2014charged}, \cite{bhar2016anisotropic}, \cite{sharma2007class}, \cite{komathiraj2007analytical}, \cite{bhar2015singularity}, \cite{bhar2017compact}, \cite{bhar2016modelling}, \cite{thomas2017anisotropic}. \cite{das2019new} have studied the metric potential in the form $ B^{2}_{0}(r)=\frac{1}{(1-\frac{r^2}{R^2})^4} $ and have shown that it can represent a viable model of compact objects. \cite{das2021modeling} have studied the metric potential in the form $ B^{2}_{0}(r)=\frac{1}{(1-\frac{r^2}{R^2})^6} $ and have shown that it can represent a viable model of compact star  4U1820-30. 

To develop a physically reasonable model of the stellar configuration, we assume that the metric potential $ g_{rr} $  co-efficient is expressed as $ B^2 $ given by
\begin{equation}
	B^{2}_{0}(r)=\frac{1}{(1-\frac{r^2}{R^2})^n},
\end{equation}
where  $ n > 0 $ is any real number. By selecting this metric potential, the function $ B^{2}_{0}(r)  $ is guaranteed to be finite, continuous and well-defined within the range of stellar interiors. Also $ B^{2}_{0}(r)=1 $ for $ r=0 $  ensures that it is
finite at the center. Again, the metric is regular at the center
since $(B^{2}_{0}(r))'_{r=0} =0. $ 
\\ With the choice of $B^{2}_{0}(r) , $ equation (\ref{e5}) reduces to
\begin{equation}\label{e6}
	\Delta(r)=\frac{1}{\left( 1-\frac{r^2}{R^2}\right) ^{-n}}\left[ \frac{A_{0}^{''}}{A_{0}}-\frac{A_{0}^{'}}{A_{0}}\left( \frac{1}{r}+\frac{n\frac{r^2}{R^2}}{(1-\frac{r^2}{R^2})}\right) -\frac{n}{(1-\frac{r^2}{R^2})}-\frac{1}{r^2}+\frac{(1-\frac{r^2}{R^2})^{-n}}{r^2}\right],
\end{equation}
Rearranging equation (\ref{e6})  we get,
\begin{equation}\label{e7}
	\frac{A_{0}^{''}}{A_{0}}-\frac{A_{0}^{'}}{A_{0}}\left( \frac{1}{r}+\frac{n\frac{r^2}{R^2}}{(1-\frac{r^2}{R^2})}\right) -\frac{n}{(1-\frac{r^2}{R^2})}-\frac{1}{r^2}+\frac{(1-\frac{r^2}{R^2})^{-n}}{r^2}=\Delta(r)\left( 1-\frac{r^2}{R^2}\right) ^{-n},
\end{equation}
We choose $ \Delta(r) $ to solve equation (\ref{e7}) as
\begin{equation}\label{delta}
	\Delta(r)=\left( 1-\frac{r^2}{R^2}\right) ^{n}\left( \frac{-n}{(1-\frac{r^2}{R^2})}-\frac{1}{r^2}+\frac{(1-\frac{r^2}{R^2})^{-n}}{r^2}\right). 
\end{equation}
The above choice for anisotropy is physically reasonable,
as at the center $ (r=0) $ anisotropy vanishes as expected.
Substituting equation (\ref{delta}) in (\ref{e7}), we obtain,
\begin{equation}\label{e9}
	\frac{A_{0}^{''}}{A_{0}}-\frac{A_{0}^{'}}{A_{0}}\left( \frac{1}{r}+\frac{n\frac{r^2}{R^2}}{(1-\frac{r^2}{R^2})}\right)=0.
\end{equation}
We obtaine a simple solution of the equation (\ref{e9})
\begin{equation}
	A_{0}(r)=\frac{C\left( R^2-r^2\right) ^{1-\frac{n}{2}}}{n-2}+D.
\end{equation}
The interior spacetime metric takes the form
\begin{equation}
	ds^2 = -\left( \frac{C\left( R^2-r^2\right) ^{1-\frac{n}{2}}}{n-2} + D \right) ^{2} dt^2 + \frac{1}{(1-\frac{r^2}{R^2})^n} dr^2 + r^2(d\theta^2 +sin^2 \theta d\phi^2).
\end{equation}
The matter density, radial pressure and transverse pressure takes the form,
\begin{equation}
	8\pi\rho=-\frac{-1+\left( 1-\frac{r^2}{R^2}\right) ^{n}}{r^2}+\frac{2n\left( 1-\frac{r^2}{R^2}\right) ^{-1-n}}{R^2},
\end{equation}
\begin{equation}
	8\pi p_{r}=\frac{-1+\left( 1-\frac{r^2}{R^2}\right) ^{n}}{r^2}-\frac{2C(-2+n)\left( 1-\frac{r^2}{R^2}\right) ^{n}}{C(r^2-R^2)-D(-2+n)(-r^2+R^2)^{n/2}},
\end{equation}
\begin{equation}
	8\pi p_{\perp}=-\frac{\left( 1-\frac{r^2}{R^2}\right) ^{n}\left( C(-4+n)(r^2-R^2)+D(-2+n)n(-r^2+R^2)^{n/2}\right)}{(r^2-R^2)\left( C(r^2-R^2)-D(-2+n)(-r^2+R^2)^{n/2}\right) },
\end{equation}
\\ Where C and D are integration constants, which will be determined using boundary condition.
\section{Exterior space-time and boundary conditions}
At the boundary of the star $ r = b $, we match the interior
metric (\ref{e1}) with the Schwarzschild exterior spacetime metric.

\begin{equation}\label{e8}
	ds^{2}=-\left(1-\frac{2m}{r}\right)dt^{2}+\left(1-\frac{2m}{r}\right)^{-1}dr^{2} + r^2(d\theta^2+sin^2\theta.d\phi^2),
\end{equation} 
which leads to
\begin{equation}
	A^{2}_{0}(r)=\left(1-\frac{2m}{b}\right),
\end{equation}
\begin{equation}
	B^{2}_{0}(r)=\left(1-\frac{2m}{b}\right)^{-1},
\end{equation}
At the boundary of stars $p_{r}(r = b) = 0 $ which gives,

\begin{equation}
	R = \sqrt{\frac{b^2}{1-\sqrt[n]{1-\frac{2m}{b}}}},
\end{equation}
\begin{equation}
	C=\frac{R^{n}\left( -1+\left( 1-\frac{b^2}{R^2}\right) ^{n}\right) }{-2b^{2}},
\end{equation}
\begin{equation}
	D= \left( 1-\frac{b^2}{R^2}\right) ^{n/2}+\frac{R^{n}\left( -1+\left( 1-\frac{b^2}{R^2}\right) ^{n}\right)(R^2-b^2)^{1-\frac{n}{2}} }{2(n-2)b^{2}},
\end{equation}
When the values of C and D are substituted in $ p_{r}$ and  $ p_{\perp} $ we get,

\begin{equation*}
	8\pi p_{r}=\frac{-1+\left( 1-\frac{r^2}{R^2}\right) ^{n}}{r^2} +
\end{equation*}
\begin{equation}
	\frac{\left( 1-\left( 1-\frac{b^2}{R^2}\right) ^{n}\right) \left( 1-\frac{r^2}{R^2}\right) ^{n} R^{n}(R^{2}-r^{2})^{\frac{-n}{2}}}{b^{2}\left( \left( 1-\frac{r^2}{R^2}\right) ^{n/2} +\frac{\left( -1+\left( 1-\frac{b^2}{R^2}\right) ^{n}\right) R^{n}(R^{2}-b^{2})^{1-\frac{n}{2}}}{2b^{2}(n-2)}-\frac{\left( -1+\left( 1-\frac{b^2}{R^2}\right) ^{n}\right) R^{n}(R^{2}-r^{2})^{1-\frac{n}{2}}}{2b^{2}(n-2)}\right)},
\end{equation}
\begin{equation}
	8\pi p_{\perp}=8\pi p_{r}+\Delta(r),
\end{equation}

\begin{table}[h]
	\caption{The numerical values of the strong energy condtion at center as well as surface, redshift at surface and adiabatic Index at surface for the compact star  PSR J1903+327.}
	\label{tab:1}
	\begin{tabular}{lllll}
		\hline\noalign{\smallskip}
		\textbf{n} & 
		{$ \mathbf{ \rho - p_{r} - 2p_{\perp}}_{(r=0)} $} & {$ \mathbf{\rho-p_{r}-2p_{\perp}}_{(r=b)} $} &    {$ \mathbf{ Z_{(r=b)}} $} &  {$ \mathbf{\Gamma_{(r=0)}}$} 
		\\	&   \textbf{(MeV fm{$\mathbf{^{-3}}$})} & \textbf{(MeV fm{$\mathbf{^{-3}}$})} &    \textbf{(Redshift)} & \textbf{(Adiabatic }   \\
		& \textbf{}	&  \textbf{} & \textbf{} & \textbf{ Index)}   \\
		\noalign{\smallskip}\hline\noalign{\smallskip}
		\textbf{$ n = 4 $} 	  & 377.861  & 346.899 &    0.44166   & 3.59 \\
		\textbf{$ n = 6 $} 	    & 418.23   & 330.826 &   0.44166   & 3.15\\
		\textbf{$ n = 10 $} 	  & 451.728    & 318.434  & 0.44166 & 2.85\\
		\textbf{$ n = 15 $}    & 468.892   & 312.388 &  0.44166   & 2.71 \\
		\textbf{$ n = 20 $}    & 477.58   & 309.402 &   0.44166   & 2.65 \\
		\textbf{$ n = 50 $}    & 493.40   & 304.089  & 0.44166    & 2.54\\
		\textbf{$ n = 70 $}    & 496.439   & 303.085  &   0.44166    & 2.52\\
		\noalign{\smallskip}\hline
	\end{tabular} 
\end{table}
Next section contains physical analysis.

\begin{table}[h]
	\caption{The numerical values of the $ \frac{dp_{r}}{d\rho} $ at center as well as surface and $ \frac{dp_{\perp}}{d\rho} $ at center as well as surface for the compact star  PSR J1903+327.}
	\label{tab:2}
	\begin{tabular}{llllll}
		\hline\noalign{\smallskip}
		\textbf{n} &  {$ \mathbf{\frac{dp_{r}}{d\rho}_{(r=0)}} $} & {$ \mathbf{\frac{dp_{\perp}}{d\rho}_{(r=0)}} $}  & {$ \mathbf{\frac{dp_{r}}{d\rho}_{(r=b)}} $} & {$ \mathbf{\frac{dp_{\perp}}{d\rho}_{(r=b)}} $} & {$ \mathbf{(\nu^{2}_{t}-\nu^{2}_{r})_{(r=b)}} $}\\
		&   \textbf{} & \textbf{} & \textbf{}   \\
		\noalign{\smallskip}\hline\noalign{\smallskip}
		\textbf{$ n = 4 $} 	 &  0.462272  & 0.262272  & 0.360596 & 0.19339 & -0.1672 \\
		\textbf{$ n = 6 $} 	  &  0.3719  & 0.1719   &  0.3130 & 0.1517 & -0.1613 \\
		\textbf{$ n = 10 $}  &  0.31227  & 0.1122   &  0.2806 & 0.1338 & -0.1468 \\
		\textbf{$ n = 15 $}   &  0.2858  & 0.0858   &  0.2658 & 0.1236  & -0.1434 \\
		\textbf{$ n = 20 $}   &  0.2733  & 0.0733   &  0.2587 & 0.1188 &-0.1422 \\
		\textbf{$ n = 50 $}    &  0.2520 & 0.052   &  0.2465 & 0.1108 & -0.1357 \\
		\textbf{$ n = 70 $}    &  0.2480  & 0.0480   &  0.2443 & 0.1093 & -0.135 \\
		\noalign{\smallskip}\hline
	\end{tabular} 
\end{table}
\begin{table}[h]
	\caption{The values of the curvature parameter R for the compact star PSR J1903+327  whose observed mass and radius is
		given by $ 1.66^{+ 0.021}_{- 0.021} M_\odot $ and $ 9.438^{+0.03}_{-0.03}  km $ respectively \cite{gangopadhyay2013strange}.}
	\label{tab:3}
	\begin{tabular}{lllllll}
		\hline\noalign{\smallskip}
		\textbf{Star} & {$ \textbf{ Estimated} $} & {$ \textbf{Estimated} $} &   {$ \textbf{n} $} & {$ \textbf{R} $} 		\\	
		&   \textbf{$ Mass M_\odot  $} & \textbf{Radius} &    &\textbf{ Km }\\
		
		\noalign{\smallskip}\hline\noalign{\smallskip}
		$ PSR J1903+327 $ & 1.66 & 9.438	& \textbf{$ n = 4 $}  & 23.085  		  \\ 
		&  &   & \textbf{$ n = 6 $}  & 27.856  \\
		&  &   & \textbf{$ n = 10 $}  & 35.533  \\
		&  &   & \textbf{$ n = 15 $}  & 43.257 \\
		&  &   & \textbf{$ n = 20 $}  & 49.796 \\
		&  &   & \textbf{$ n = 50 $}  & 78.309 \\
		&  &   & \textbf{$ n = 70 $}  & 92.560 \\
		\noalign{\smallskip}\hline
	\end{tabular} 
\end{table}

\section{Physical analysis}
1. The gravitational potentials in this model satisfy,
\\$ A^{2}_{0}(r=0)=\frac{C\left( R^2\right) ^{1-\frac{n}{2}}}{n-2}+D $ = constant,
\\ $ B^{2}_{0}(r=0)=1 $
i.e., finite at the center (r = 0) of the stellar configuration. One can also
easily check that $ (A^{2}_{0}(r))^{'}_{r=0}=(B^{2}_{0}(r))^{'}_{r=0}=0 $. These results indicate that the metric is regular in the centre and behaves well throughout the stellar interior.

2.  The central density, central radial pressure and central tangential pressure in this case are:
\begin{equation*}
	\rho(0)=\frac{3n}{R^2},
\end{equation*}
\begin{equation*}
	p_{r}(0)=\frac{-n}{R^2}+\frac{2C(n-2)}{-C(R^2)-D(n-2)(R^2)^{\frac{n}{2}}},
\end{equation*}
\begin{equation*}
	p_{\perp}(0)=\frac{-n}{R^2}+\frac{2C(n-2)}{-C(R^2)-D(n-2)(R^2)^{\frac{n}{2}}}.
\end{equation*}
Note that the density is always positive as R is a positive
quantity. The radial pressure and tangential pressure at
the centre are equal which means pressure anisotropy
vanishes at the center. The radial and tangential pressures
at the center will be non-negative if one chooses the model
parameters satisfying the conditions $ n > 2. $ In Fig.(\ref{Figure:1}) we have shown the variation of density for $0\leq r \leq 9.438 $. It is clear from the graph that the density is a decreasing function of $r$. In Fig.(\ref{Figure:2}) and Fig.(\ref{Figure:3}), we have shown the variation of radial and transverse pressures throughout the star. It can be seen that both pressures are decreasing radially outward.

\begin{figure}[ht]
	\centering
	\includegraphics[scale = 1]{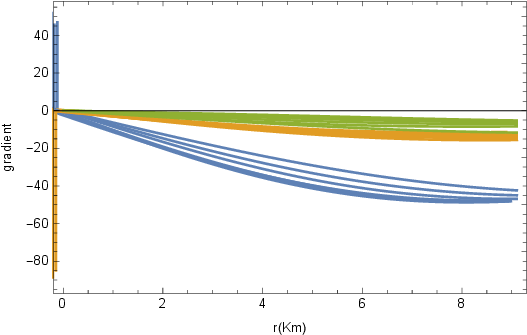}
	\caption{Variation of a Gradients $ \frac{d\rho}{dr}(Blue) $,$ \frac{dp_{r}}{dr}(Orange) $ and $ \frac{dp_{\perp}}{dr}(Green) $ for the compact star PSR J1903+327.
		\label{Figure:14}}
\end{figure}

3. The gradient of density, radial pressure and tangential pressure are all negative inside the stellar body, as shown in Fig.(\ref{Figure:14}), which graphically represents the compact star PSR J1903+327 for the different values of 'n' in the following section.

4. In this model the speed of sound is less than 1 in the
interior of the star, i.e.,$ 0\leq\dfrac{dp_{r}}{d\rho}\leq 1 $ , $ 0\leq\dfrac{dp_{\perp}}{d\rho}\leq 1  $
which has been shown graphically in the next section. In Fig.(\ref{Figure:4}) and Fig.(\ref{Figure:5}), we have displayed the variation of $\frac{dp_{r}}{d\rho}$ and $\frac{dp_{\perp}}{d\rho}$ against $r$. Both quantities satisfy the restriction $ 0<\frac{dp_{r}}{d\rho}<1 $ and $ 0<\frac{dp_{\perp}}{d\rho}<1 $ indicating that the sound speed is less than the speed of light throughout the star. Table(\ref{tab:2}) shows the values of $ \frac{dp_{r}}{d\rho}  $ and $ \frac{dp_{\perp}}{d\rho}   $ at a center as well as the surface of the star.

5.  Energy condition: 
\\The following are the most crucial requirements for our model to be a physically plausible, i.e. the null energy condition (NEC), weak energy condition (WEC), dominent energy condition (DEC) and strong energy condition (SEC). If all of the following inequalities exist at the same time, these energy criteria are satisfied:
\begin{equation}
	NEC:\rho\ge 0,  
\end{equation}

\begin{equation}
	WEC:\rho \ge 0 , \rho+p_{r}\ge 0 , \rho+p_{\perp}\ge 0,
\end{equation}

\begin{equation}
	DEC: \rho\ge \mid p_{r} \mid, \rho\ge \mid p_{\perp} \mid,
\end{equation}

\begin{equation}
	SEC: \rho - p_{r} - 2p_{\perp}\ge0.
\end{equation}

\begin{figure}[ht]
	\centering
	\includegraphics[scale = 1]{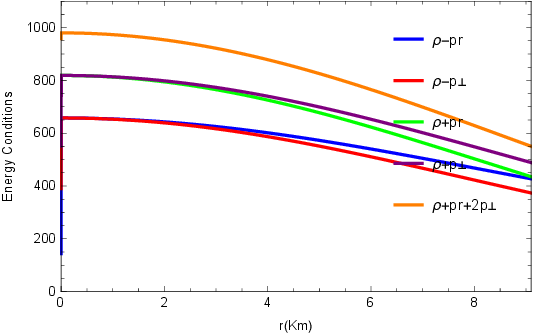}
	\caption{All the energy conditions are plotted against $ r $ inside the stellar interior for the compact star PSR J1903+327 for ‘n = 70’ mentioned in the figures.  
		\label{Figure:13}}
\end{figure}

The Fig(\ref{Figure:13}) demonstrate that our model of a compact star satisfies all of the energy requirements for the value of 'n=70' of the compact star PSR J1903+327. Fig(\ref{Figure:7}) indicates that the strong energy condition $\rho-p_{r}-2p_{\perp}>0$ is satisfied throughout the distribution. Table(\ref{tab:1}) shows the values of $ \rho-p_{r}-2p_{\perp}  $ at a center as well as the surface of the star.

\begin{figure}[ht]
	\centering
	\includegraphics[scale = 1]{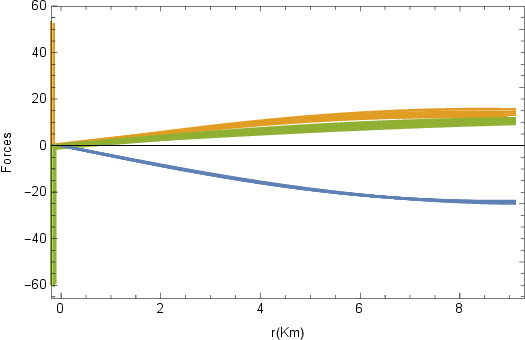}
	\caption{Variation of three forces like Gravitational Force(Blue), Hydrostatic Force(Orange) and Anisotropic Force(Green) for the compact star PSR J1903+327. 
		\label{Figure:12}}
\end{figure}

6. Stability under three forces:

 The equation can be used to describe the stability of our current model under three forces: gravity force, hydrostatic force and anisotropic force.
\begin{equation}\label{TOV}
	-\frac{M_{G}(r)(\rho+p_{r})}{r^2}\frac{B}{A}-\dfrac{dp_{r}}{dr}+\frac{2}{r}(p_{\perp}-p_{r})=0,
\end{equation}
The Tolman-Whittaker formula and the Einstein's field equations may be used to calculate $ M_{G}(r) $, which stands for the gravitational mass inside the radius r, and is defined by
\begin{equation}
	M_{G}(r)=r^2\frac{A'}{B}.
\end{equation}
Equation (\ref{TOV}) changes to,
\begin{equation}\label{TOV2}
	F_{g}+F_{h}+F_{a}=0,
\end{equation}
\begin{equation}
	F_{g}=-\frac{2C(n-2)r(1-\frac{r^2}{R^2})^n(2C(n-1)(r^2-R^2)-Dn(n-2)(R^2-r^2)^{\frac{n}{2}})}{(r^2-R^2)(\Phi)^2},
\end{equation}
\begin{equation*}
	F_{h}=-\frac{2(-1+\frac{r^2}{R^2})^n}{r^3}-\frac{2n(1-\frac{r^2}{R^2})^{n-1}}{rR^2}+\frac{4Cnr(n-2)(1-\frac{r^2}{R^2})^{n-1}}{R^2(\Phi)^2}
\end{equation*}
\begin{equation}
	+\frac{2C(n-2)(1-\frac{r^2}{R^2})^{n}(2Cr+Dnr(n-2)(R^2-r^2)^{(\frac{n-2}{2})})}{(\Phi)^2},
\end{equation}
\begin{equation}
	F_{a}=\frac{2\left( r^2\left( 1-(1-\frac{r^2}{R^2})^{n}+(1-\frac{r^2}{R^2})^{n}\right) +\left( -1+(1-\frac{r^2}{R^2})^{n}\right) R^2\right) }{r^5-r^3R^2},
\end{equation}
where
\begin{equation*}
	\Phi=C(r^2-R^2)-D(n-2)(R^2-r^2)^{\frac{n}{2}}.
\end{equation*}
Fig.(\ref{Figure:12}) shows the graphical representation of three distinct forces for the compact star PSR J1903+327 with the different values 'n'. According to the graphs, The gravitational force is a net negative force that predominates in nature. Hydrostatic and anisotropic forces work together to balance this force and keep the system in equilibrium.

\begin{figure}[ht]
	\centering
	\includegraphics[scale = 1]{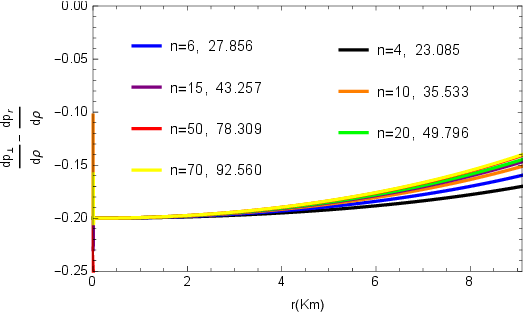}
	\caption{Graph of $ v^{2}_{t}-v^{2}_{r} $. 
		\label{Figure:11}}
\end{figure}

7. Causality condition and method of cracking:

The causality criterion states that a physically plausible model's radial  sound velocity $ v^{2}_{r} $ and transverse sound velocity  $ v^{2}_{\perp} $ must fall within the interval  [0, 1]. Where the definitions of the radial $ v^{2}_{r} $ and transverse $ v^{2}_{\perp} $ velocities of sound are
\begin{equation}
	v^{2}_{r}=\frac{p_{r}^{\prime}}{\rho^{\prime}} ,\;\;\;\ v^{2}_{\perp}=\frac{p_{\perp}^{\prime}}{\rho^{\prime}},
\end{equation}
we can easily derive $ \rho^{\prime}  $ , $ p_{r}^{\prime}  $ and $ p_{\perp}^{\prime} $ using the expression $ \rho $ , $ p_{r} $ and $ p_{\perp}. $
Fig(\ref{Figure:4}) and Fig(\ref{Figure:5}) displays the $ v^{2}_{r} $ and $ v^{2}_{\perp} $ profiles. The figures make it evident that both $ v^{2}_{r} $ and $ v^{2}_{\perp} $ fall within an acceptable range. Therefore, it may be said that, the causality criterion is effectively satisfied. \cite{herrera1992cracking} suggested the technique of ``cracking" for the stability of a compact star model and utilising this technique. \cite{abreu2007sound}  proposed that, for a possibly stable configuration, $ v^{2}_{\perp}-v^{2}_{r} < 0 $, the stability factor is negative, as can be seen in Fig(\ref{Figure:11}). With this, We conclude that our model may be stable everywhere in the stellar interior. 

8. Adiabatic index:

\cite{bondi1964contraction} investigated whether a Newtonian isotropic sphere will be in equilibrium for a specific stellar configuration, if the adiabatic index $ (\Gamma) > 4/3  $ and it is adapted for a relativistic anisotropic fluid sphere.
Based on these findings, it can be determined that an anisotropic star configuration's stability depends on the adiabatic index $ \Gamma $ by,
\begin{equation*}
	\Gamma_{r}=\frac{\rho+p_{r}}{p_{r}}\frac{dp_{r}}{d\rho},
\end{equation*}
\begin{equation}
	= \frac{2\left( 1-\frac{r^2}{R^2}\right) ^{n}\left(- \frac{n}{r^2-R^2}-\frac{C(n-2)}{\Phi}\right)}{\frac{-1+( 1-\frac{r^2}{R^2}) ^{n}}{r^2}-\frac{2C(n-2)( 1-\frac{r^2}{R^2}) ^{n}}{\Phi}}v^{2}_{r}.
\end{equation}
For various values of $ 'n' $, the profile of the adiabatic index ($ \Gamma _{r} $) for our current model is shown in Fig(\ref{Figure:8}). The graph shows that the radial adiabatic index profile is a monotonic increasing function of r and that $ \Gamma =\frac{\rho+p_{r}}{p_{r}}\frac{dp_{r}}{d\rho}>\frac{4}{3} $ everywhere in the star arrangement, satisfying the stability requirement. Adiabatic index indicates that the condition $\Gamma>\frac{4}{3}$ is satisfied in the region $ 0\leq r\leq 9.438 $. Table(\ref{tab:1}) shows the values of $ \Gamma_{r}  $ at a center of the star.

\section{Conclusion}

\begin{figure}[ht]
	\centering
	\includegraphics[scale = 1]{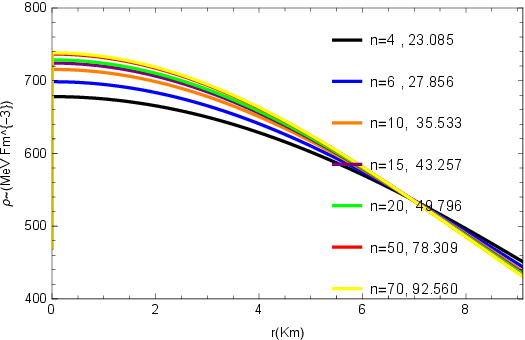} 
	\caption{Variation of density against radial variable $r$. 
		\label{Figure:1}}
\end{figure}

\begin{figure}[ht]
	\centering
	\includegraphics[scale = 1]{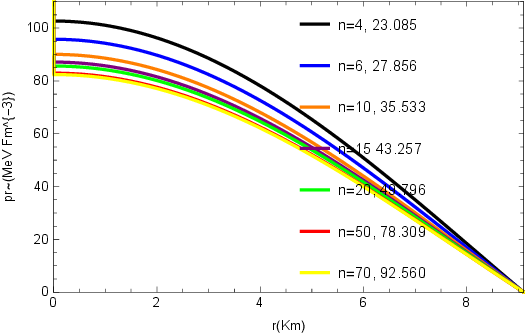}
	\caption{Variation of radial pressures against radial variable $r$.
		\label{Figure:2}}
\end{figure}

\begin{figure}[ht]
	\centering
	\includegraphics[scale = 1]{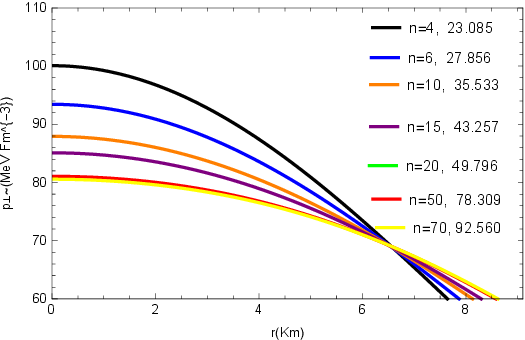}
	\caption{Variation of transverse pressures against radial variable $r$ 
		\label{Figure:3}}
\end{figure}

\begin{figure}[ht]
	\centering
	\includegraphics[scale = 1]{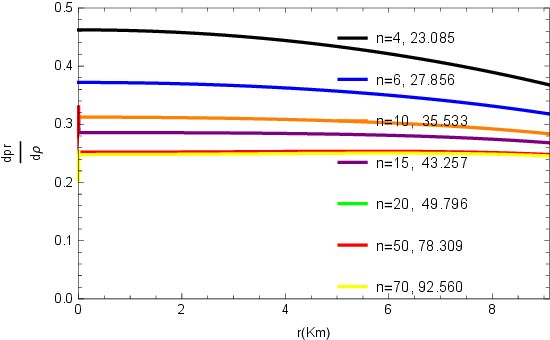}
	\caption{Variation of $ \frac{dp_r}{d\rho} $ against radial variable $r$. 
		\label{Figure:4}}
\end{figure}

\begin{figure}[ht]
	\centering
	\includegraphics[scale = 1]{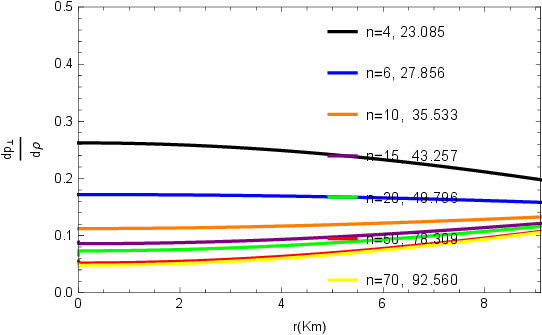}
	\caption{Variation of $ \frac{dp_\perp}{d\rho} $ against radial variable $r$. 
		\label{Figure:5}}
\end{figure}

\begin{figure}[ht]
	\centering
	\includegraphics[scale=1]{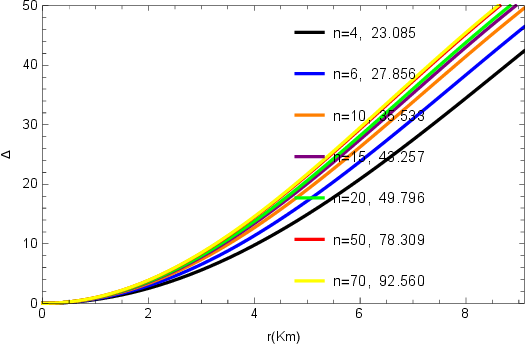}
	\caption{Variation of anisotropies against radial variable $r$. 
		\label{Figure:6}}
\end{figure}

\begin{figure}[ht]
	\centering
	\includegraphics[scale = 1]{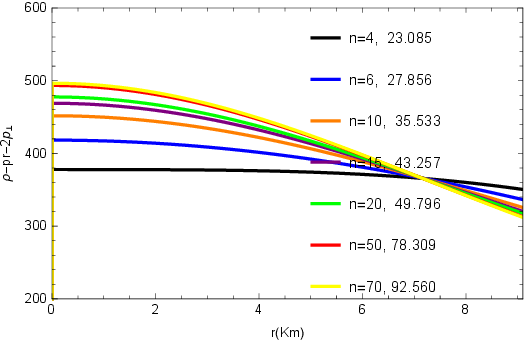}
	\caption{Variation of strong energy condition against radial variable $ r $. 
		\label{Figure:7}}
\end{figure}

\begin{figure}[ht]
	\centering
	\includegraphics[scale = 1]{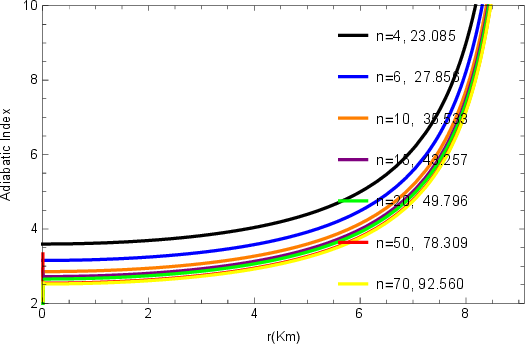}
	\caption{Variation of Adiabatic Index against radial variable $r$. 
		\label{Figure:8}}
\end{figure}

By making the assumption of a pressure anisotropy and a metric potential that is physically plausible, the current work offers a new generalised model of compact stars.
To set the values of the various constants, we compared our interior solution to the exterior Schwarzschild line component at the boundary. For the compact star  PSR J1903+327  with masses and radii $ M = 1.66 M{\odot}  $ and  b = 9.438 km  respectively,  we have determined the values of R, C and D from the boundary conditions for various values of the dimensionless parameter ''n". For the compact star PSR J1903+327, different values of "n" satisfy all physical plausibility conditions such as matter density($ \rho $), radial pressure$ (p_{r}) $ and transverse pressure$(p_{\perp}) $, sound speed of radial and transverse ($ \frac{dp_{r}}{d\rho} $ and $ \frac{dp_{\perp}}{d\rho} $), adiabatic Index($ \Gamma $), surface redshift(Z) and strong energy condition$ (\rho-p_{r}-2p_{\perp}).$ We can observe from the Fig.(\ref{Figure:1}), (\ref{Figure:2}), Fig.(\ref{Figure:3}), (\ref{Figure:4}), Fig.(\ref{Figure:5}), (\ref{Figure:6}), Fig.(\ref{Figure:7}), (\ref{Figure:8}) that these are all positive and monotonically decreasing functions of r. Hence the model is suitable to describe PSR J1903+327.


\bibliography{sn-jnl.bib}



\begin{figure}[ht]
	\centering
	\includegraphics[scale = 1]{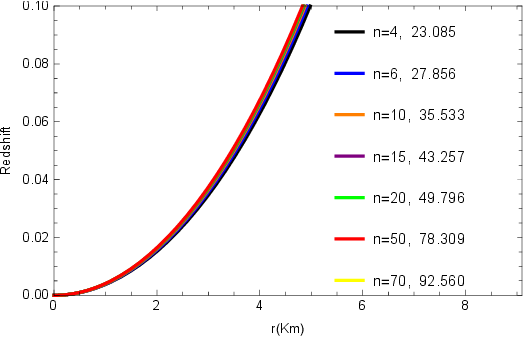}
	\caption{Variation of Redshift against radial variable $r$. 
		\label{Figure:9}}
\end{figure}

\begin{figure}[ht]
	\centering
	\includegraphics[scale = 1]{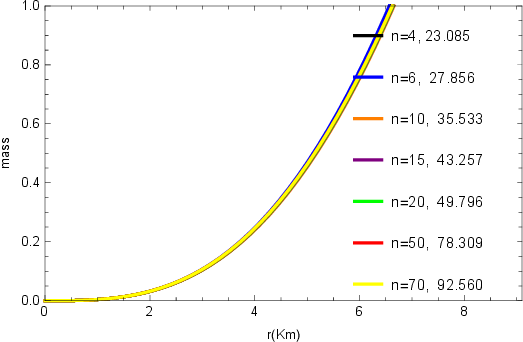}
	\caption{Variation of Mass against radial variable $r$. 
		\label{Figure:10}}
\end{figure}


\end{document}